\documentclass[12pt,draft,onecolumn]{IEEEtran}
\usepackage{amsmath,amssymb,threeparttable,placeins,makecell,stfloats,cite}
\usepackage[dvips]{graphicx,epstopdf}
\usepackage[usenames]{color}
\usepackage{amsfonts}
\usepackage{latexsym}
\usepackage{subfigure}
\usepackage[justification=centering]{caption}
\usepackage{floatrow}
\newtheorem{theorem}{{{\textit{Theorem}}}}

\newtheorem{lemma}{{{\textit{Lemma}}}}
\newtheorem{corollary}{{{{\textit{Corollary}}}}}
\newtheorem{property}{{{\textit{Property}}}}
\newtheorem{definition}{{{\textit{Definition}}}}

\newtheorem{remark}{{{\textit{Remark}}}}

\newtheorem{example}{{{\textit{Example}}}}

\newtheorem{construction}{{{\textit{Construction}}}}

\begin{document}
	
\title{Asymptotically Optimal and Near-optimal Aperiodic Quasi-Complementary Sequence Sets Based on Florentine Rectangles}
\author{Avik Ranjan Adhikary,~\IEEEmembership{Member,~IEEE},
	Yanghe Feng,
    Zhengchun Zhou,~\IEEEmembership{Member,~IEEE},
    and Pingzhi Fan, ~\IEEEmembership{Fellow,~IEEE}.
%%	% <-this % stops a space
  \thanks{ 
%  	This work was supported in part by  National Natural Science Foundation of China (No. 61661146003, No. 61672028 and No.61731017), in part by the Application Fundamental Research Plan Project of Sichuan
%  	Province under Grant 2018JY0046, and in part by 111 Project (No.111-2-14).
%  	
%  	
%  	   Manuscript received Nov. 07, 2019; revised Jan. 06, 2020; accepted Feb. 26, 2020. The editor coordinating the review of this manuscript and approving it for publication was Prof. Rafael Schaefer.
%  
%  (\textit{Corresponding author: Avik Ranjan Adhikary.})
%  
%  	
Avik Ranjan Adhikary and Zhengchun Zhou are with School of Mathematics, Southwest Jiaotong University,
Chengdu, 610031, China. E-mail: {\tt Avik.Adhikary@ieee.org, zzc@swjtu.edu.cn}.
Yanghe Feng is with the College of Systems Engineering, National University of Defense Technology,
Changsha, 410073, China. E-mail: {\tt fengyanghe@nudt.edu.cn}. Pingzhi Fan is with Institute of Mobile Communications, Southwest Jiaotong University,
Chengdu, 610031, China. E-mail: {\tt p.fan@ieee.org}. 
}
}
\maketitle

\begin{abstract}
	
Quasi-complementary sequence sets (QCSSs) can be seen as a generalized version of complete complementary codes (CCCs), which enables multicarrier communication systems to support more users. The contribution of this work is two-fold. First, we propose a systematic construction of Florentine rectangles. Secondly, we propose several sets of CCCs and QCSS, using Florentine rectangles. The CCCs and QCSS are constructed over $\mathbb{Z}_N$, where $N\geq2$ is any integer. The cross-correlation magnitude of any two of the constructed CCCs is upper bounded by $N$. By combining the proposed CCCs, we propose asymptotically optimal and near-optimal QCSSs with new parameters. This solves a long-standing problem, of designing asymptotically optimal aperiodic QCSS over $\mathbb{Z}_N$, where $N$ is any integer.
\end{abstract}


\section{INTRODUCTION}
\IEEEPARstart{G}{olay} complementary pairs (GCPs) were proposed by M. J. Golay in 1951 in his work on multislit spectrometry \cite{golay}. Golay complementary pairs are a pair of sequences whose aperiodic autocorrelations sum up to zero at each non-zero time shift \cite{golay,golay1}. In 1972, Tseng and Liu extended the concept from complementary pairs to complementary sets (CS) \cite{Tseng}. In a remarkable work in 1988, Suehiro and Hatori proposed N-shift cross orthogonal sequences \cite{suehiro}, which were later termed as perfect complementary sequence sets (PCSS). A set of $K$ mutually orthogonal CS, where each CS consists of $M$ sequences (also known as sub-carriers), each of length $N$, is called a $(K,M,N)$- PCSS. PCSSs are also known as mutually orthogonal Golay complementary sets (MOGCS) \cite{rathina}. Owing to their ideal correlation properties, PCSSs have been widely used in multi-carrier code division multiple access (MC-CDMA) systems for the reduction of the peak-to-average power ratio (PAPR) \cite{Davis-1999}, channel estimation \cite{Spasojevic-2001}, \cite{Wang-2007}, etc. One of the main drawbacks of PCSS with $M$ sub-carriers is that, when used in MC-CDMA systems, it can support at most $M$ users \cite{zilongccc,chen1}.

%%%%%%%%%%%%%%%%%%%%%%%%%%%%%%%%%%%%%%%%%%%%%%%%%%%%%%%%%%%5
%Perfect complementary sequence set (PCSS) refers to a set of  two-dimensional matrices with non-trivial autocorrelations and cross-correlations summing to zero for any non-zero time-shift. Such sequence sets have wide applications due to their ideal correlation properties, such as reducing the peak-to-average power ratio (PAPR) \cite{Davis-1999}, channel estimation \cite{Spasojevic-2001}, \cite{Wang-2007}, RADAR waveform design \cite{Pezeshki-2008}, etc. In typical applications of complementary set, each constituent
%sequence should be sent out in a separate channel and
%thus each complementary set is transmitted in a “multichannel” system. When a multi-carrier code division multiple access (MC-CDMA)
%transmission of complementary sets is considered,
%due to the mutual orthogonality of the complementary sets of the PCSS, a PCSS-MC-CDMA
%with $M$ sub-carriers is capable of supporting at most $M$ users \cite{zilongccc,chen1}.

%%%%%%%%%%%%%%%%%%%%%%%%%%%%%%%%%%%%%%%%%%%%%%%%%%%%%%%%%%%%%%%

%PCSSs can support only a limited number of users in multi-carrier code division multiple access (MC-CDMA) systems since the set size (i.e. the number of users) must not be larger than the flock size (i.e. the number of channels) \cite{chen1}.

Working towards the goal of enabling MC-CDMA systems to support more users, Liu \textit{et al.} \cite{zilong111} designed low correlation zone complementary sequence sets (LCZ-CSSs) in 2011. Later in 2013, Liu \textit{et al.} designed quasi-complementary sequence sets (QCSSs) \cite{zilong13} by generalizing the concept of LCZ-CSSs. The concept of QCSS also includes Z-complementary sequence sets (ZCSSs) \cite{zhou15,zilong14,yubo14,Avik2,sarkar18,Avik5,chen17}. A QCSS of set size $K$, flock size $M$, sequence length $N$ and maximum aperiodic or periodic correlation tolerance $\delta_{\max }$, is written as $(K,M,N,\delta_{\max })$- QCSS. For QCSS, the set size $K$ denotes the number of users it can support, the flock size $M$ denotes the number of sub-carriers. When $K\leq M$ and the periodic or aperiodic correlation tolerance $\delta_{\max}=0$, QCSS becomes PCSS. When $K=M$, a PCSS is called a complete complementary code (CCC)  \cite{zilong13}.

The first correlation lower bound of sequences was given by Welch \cite{welch} in 1974. Later, in a series of works in 2011, 2014 and 2017 \cite{zilong11,zilong14_1,zilong17}, Liu \textit{et al.} derived some special conditions for aperiodic QCSSs and further tightened the lower bound. We call a QCSS optimal if $\delta_{\max}$ achieves these lower bounds.

%In 1974, Welch \cite{welch} derived a lower bound for the correlation of sequences. Later in 2011, Liu \textit{et al.} \cite{zilong11,zilong14_1} tightened the lower bound for aperiodic QCSSs when the set size $K\geq 3M$. Liu \textit{et al.} \cite{zilong17} further tightened the lower bound for the cases of aperiodic QCSSs, when $\frac{\lfloor\frac{\pi^2M}{4}\rfloor}{M}<\frac{K}{M}<3+\frac{1}{M}$ for sufficiently large $N$.

Systematic constructions of optimal QCSS, both periodic and aperiodic, for various parameters remains very challenging till date. In \cite{zilong13}, Liu \textit{et al.} designed periodic optimal QCSS by using Signer difference sets. Utilizing the properties of the difference sets and almost difference sets Li \textit{et al.} gave systematic framework to construct periodic optimal and near-optimal QCSSs in \cite{Li18,Li19}. In \cite{Li19_1} and \cite{Li19_2}, Li \textit{et al.} proposed periodic QCSSs using characters over finite fields. Recently, in 2019, asymptotically optimal aperiodic QCSSs were first proposed by Li \textit{et al.} \cite{Li19_3}, based on low-correlation CSSs over the alphabet $\mathbb{Z}_N$, where $N$ is a prime integer or power of a prime integer. In \cite{Li19_4}, Li \textit{et al.} designed a systematic framework to construct aperiodic asymptotically optimal QCSSs using several sets of CCCs having prime length sequences over the alphabet $\mathbb{Z}_N$, where $N$ is a prime integer. Recently in 2020, Zhou \textit{et al.} \cite{zhou2020} proposed a general construction of QCSS over $\mathbb{Z}_N$, for any odd integer $N\geq 3$. Asymptotically optimal aperiodic QCSSs with corresponding parameters, reported till date, are given in Table \ref{tab_intro}.

%\begin{figure*}
\begin{table}
	\small
	\resizebox{\textwidth}{!}{
		\caption{Asymptotically optimal aperiodic QCSSs.\label{tab_intro}}
		\begin{tabular}{|c|c|c|c|c|c|c|}
			\hline
			References & Set Size & Flock Size &  Sequence Length &$\delta_{\max }$ &  Alphabet&  Parameter constraint(s)   \\\hline
			\cite{Li19_4} & $t(t-1)$ & $t$ & $t$& $t$ & $\mathbb{Z}_t$ &  $t$ is an odd prime. \\
			\hline
			Theorem 1 \cite{Li19_3} & $u(u+1)$ &  $u$ & $u$&$u$ &$\mathbb{Z}_u$ & $u$ is power of a prime. \\
			\hline
			Theorem 3 \cite{Li19_3} & $u^2$ & $u$ & $u-1$&$u$ & $\mathbb{Z}_u$ & $u$ is power of a prime, and $u\geq 5$. \\
			\hline
			\cite{zhou2020} & $N(t_0-1)$ & $N$ & $N$&$N$ & $\mathbb{Z}_N$ &  \makecell{$N$ is odd, $N\geq 5$, and $t_0$ is \\the smallest prime factor of $N$} \\
			\hline
			Proposed  & $N\times F(N)$ & $N$ & $N$&$N$ & $\mathbb{Z}_N$ & $N\geq 2$ is any integer. \\
			\hline
	\end{tabular} }\\
\end{table} 

%Till date, there is no systematic constructions of asymptotically optimal and near-optimal aperiodic QCSSs over $\mathbb{Z}_N$, when $N$ is not a prime or is not a power of prime integer.
%%%%%%%%%%%%%%%%%%%%%%%%%%%%%%%%%%%%%%%%%%%%%%%%%%

Analysing closely the results of \cite{Li19_3}, \cite{Li19_4} and \cite{zhou2020} it is being observed that the number of CCC's and eventually the set size of the QCSS are small when $N$ is $3$ or have the smallest prime factor $3$. Also it has been observed in all the previous constructions \cite{Li19_3,Li19_4,zhou2020} that the optimal QCSS are designed over $\mathbb{Z}_N$, where $N$ always a prime, power of prime or an odd integer, depending on the constructions. To overcome these problems, in search of new approaches to design QCSSs over any alphabet size $N$, we propose several sets of CCCs and eventually QCSSs using Florentine rectangles. 

Florentine rectangles are extensively studied since 1989 \cite{golomb,taylor,song}. Almost all of the studies are focused on searching the existence of Florentine rectangles of given orders. Also, most of the available examples are based on computer search results. To the best of the authors' knowledge systematic constructions of Florentine rectangles are available for few particular orders of Florentine rectangles only. To construct the CCCs and eventually the QCSS, in this paper, we give a systematic construction of Florentine rectangle having a flexible order. The proposed method drastically improves the set size of the QCSS, including the cases when $N$ have the smallest prime factor $3$. The proposed construction generates multiple CCCs and eventually asymptotically optimal QCSSs over $\mathbb{Z}_N$, where $N$ is any even integer, like $N=6,10,$ etc. To the best of the authors knowledge, QCSSs over any alphabet size $N$ are not reported before. The construction uses the intrinsic structural properties of Florentine rectangles to construct these CCCs and QCSSs. Using the proposed framework, several sets of CCCs having parameters $(N,N,N)$ are proposed. Further, utilizing the constructed CCCs, we propose ($N\times F(N),N,N,N$)- QCSS, where $F(N)\times N$ Florentine rectangles exist. Since for $F(N)\geq 3$ (except when $N=2 \text{ and }3$), the proposed QCSS have set size ($K$) $\geq 3M$, flock size ($M$) $\geq 2$ and sequence length ($N$) $\geq 2$, we will check the optimality condition using the the correlation lower bound given by Liu \textit{et al.} in \cite{zilong14_1}. The cross-correlation magnitude among the CCCs is upper bounded by $N$. The optimality factor $\rho$ of the proposed QCSSs, obtained by combining the CCCs, are approximately equal to $1$, hence resulting asymptotically optimal QCSSs. For the cases when $N=2 \text{ and }3$, we use the Welch bound \cite{welch} to calculate the optimality factor $\rho$.

The rest of this paper is organized as follows. In Section II, we recall some definitions and correlation bounds related to QCSS. In Section III, we recall the definitions of Florentine rectangles and Vatican squares. We also give some systematic constructions of Florentine rectangles and Vatican squares in this section. In Section IV, we have utilised the permutations, obtained from the Florentine rectangles, to construct several sets of CCCs. In Section V, we construct QCSS by combining the several sets of CCCs. In Section VI, we have made a comparison of our construction with the previous constructions reported in the literature. Finally, we conclude our paper in Section VII.

%
%
%
%
% we introduce Florentine squares and Florentine rectangles and some correlation bounds of QCSS. In Section III, we propose a new construction of  several sets of CCCs based on Florentine rectangles. In Section IV, we obtain asymptotically optimal aperiodic QCSS by combining these CCCs into a new set. In Section V, we make a comparison of our work with the existing works in the literature. We conclude our work in Section VI.

\section{Preliminaries}
Before we begin, let us define the notations that we will be using in the paper.
\begin{itemize}
	\item  $N$ is an integer.
	\item Let the ring of integers modulo $N$ be denoted by $\mathbb{Z}_N$.
	\item $\omega_{N}=e^{\frac{2\pi i}{N}}$ is a primitive $N$-th root of unity.
	%\item $\mathbb{Z}_N^*=\{r:1\leq r< N,~\gcd(r,N)=1\}$.
%	\item $\mathbb{Z}_N^*=\{a:1\leq a<N,~\gcd(a,N)=1\}$.
%	\item $\mathbb{Z}_N^{**}=\mathbb{Z}^*_N\setminus\{1\}$.
%	\item $\phi \circ \psi$ denotes the composition of functions $\phi$ and $\psi$, i.e., $\phi \circ \psi(x)=\phi(\psi(x))$.
	\item Let a set of sequence sets be denoted by $\mathfrak{C}$.
	\item A sequence set be denoted by $\mathcal{C}$.
	\item A sequence be denoted by $C$.
	\item The complex conjugate of $x$ is denoted by $x^*$.
%	\item $\bigcup\limits_{i=0}^{n-1} C_{i}=\{C_{0}, \dots , C_{n-1}\}$ denotes the combination of $n$ set of sequence sets $C_i$.
\end{itemize}

\begin{definition}
Let $C=\left(c_{0}, c_{1}, \cdots, c_{N-1}\right)$ and $D=\left(d_{0}, d_{1}, \cdots, d_{N-1}\right)$ be two length $N$ complex-valued sequences. The aperiodic correlation function (ACF) between $C$ and $D$ is defined as
\begin{equation}
\tilde{R}_{C, D}(\tau)=\left\{\begin{array}{ll}{\sum_{t=0}^{N-1-\tau} c_{t} d_{t+\tau}^{*},} & {0 \leq \tau \leq N-1} \\ {\sum_{t=0}^{N-1+\tau} c_{t-\tau} d_{t}^{*},} & {-N+1 \leq \tau < 0}\end{array}\right. .
\end{equation}
%where $x^*$ denotes the complex conjugate of $x$.
\end{definition}

\begin{definition}
	Consider $\mathfrak{C}=\left\{\mathcal{C}^{0}, \mathcal{C}^{1}, \cdots, \mathcal{C}^{K-1}\right\}$, consisting $K$ sequence sets, each having $M$ sequences of length $N$, i.e.,
	\begin{equation}\label{eq2}
\mathcal{C}^{k}=\left[ \begin{array}{c}{C_{0}^{k}} \\ {C_{1}^{k}} \\ {\vdots} \\ {C_{M-1}^{k}}\end{array}\right]_{M \times N}, 0 \leq k \leq K-1,
	\end{equation}
	where $ C_{m}^{k}$ is the $m$-th sequence of length $N$ and is expressed as $ C_{m}^{k}=\left(c_{m, 0}^{k}, c_{m, 1}^{k}, \cdots, c_{m, N-1}^{k}\right)$, $0 \leq m \leq M-1$. The set $\mathfrak{C}$ is called a $(K,M,N, \delta_{\max })$ quasi-complementary sequence set (QCSS) if for any $\mathcal{C}^{k_1},\mathcal{C}^{k_2}\in \mathfrak{C}$, $0\leq k_1,k_2
	\leq K-1$,
	$0 \leq \tau \leq N-1, k_{1} \neq k_{2}$ or $0<\tau \leq N-1, k_{1}=k_{2}$,
	\begin{equation}
	|\tilde{R}_{\mathcal{C}^{k_1},\mathcal{C}^{k_2}}(\tau)|=\left|\sum_{m=0}^{M-1} \tilde{R}_{C_{m}^{k_{1}},
		{C_{m}^{k_{2}}}}(\tau)\right| \leq \delta_{\max },
	\end{equation}
	where
	$K$, $M$, $N$ and $\delta_{\max }$ denotes the set size, the number of sequences in each sequence set, the length of constituent sequences, and the maximum aperiodic cross-correlation magnitude of $\mathfrak{C}$, respectively. When $K = M$ and $\delta_{\max }=0$, $(K,M,N, \delta_{\max })$-QCSS transforms into $(M,M,N)$-CCC.
\end{definition}

%For a sequence set $C^k$ defined in (\ref{eq2}), let us define the shift operator $\mathcal{L}$ as follows:
%\begin{equation}
%	\mathcal{L}^\tau(C^k)=\left[ \begin{array}{c}{C_{\tau}^{k}} \\ {C_{\tau+1}^{k}} \\ {\vdots} \\ {C_{M-1}^{k}}\\C_0^k\\\vdots\\C_{\tau-1}^k\end{array}\right]_{M \times N}
%\end{equation}

We now discuss the lower bound of $\delta_{\max }$.

\begin{lemma} {\cite{welch}}\label{lem1}
	Considering aperiodic correlation, for a QCSS with set size $K$, flock size $M$, sequence length $N$ and aperiodic correlation tolerance $\delta_{\max}$, the following inequality holds
	\begin{equation}\label{wel}
	\delta_{\max } \geq M N \cdot \sqrt{\frac{\left(\frac{K}{M}-1\right)}{K(2 N-1)-1}}.
	\end{equation}
\end{lemma}

In 2014, Liu \textit{et al.} \cite{zilong14_1} proposed a tighter lower bound of $\delta_{\max}$ for aperiodic QCSS by imposing certain restrictions on the values of $K$, $M$ and $N$.

\begin{lemma} {\cite{zilong14_1}}
	For an aperiodic QCSS with set size $K$, flock size $M$, sequence length $N$ and aperiodic correlation tolerance $\delta_{\max}$, the following inequality holds
	\begin{equation}\label{eq-1}
	\delta_{\max } \geq \sqrt{MN\left(1-2\sqrt{\frac{M}{3K}}\right)},
	\end{equation}
when $K\geq 3M$, $M\geq2$ and $N \geq 2$.
\end{lemma}

In this work, when $K\geq 3M$, i.e., for $F(N)\geq 4$, a QCSS is optimal if $\delta_{\max }$ satisfies (\ref{eq-1}) with equality. Therefore, when $K\geq 3M$, the optimality factor $\rho$ is defined as follows
\begin{equation}
\rho=\frac{\delta_{\text {max}}}{\sqrt{MN\left(1-2\sqrt{\frac{M}{3K}}\right)}}.
\end{equation}
When $K\ngeq 3M$, i.e., when $F(N)< 4$, we define the optimality factor $\rho$ as
\begin{equation}
\rho=\frac{\delta_{\text {max}}}{M N \cdot \sqrt{\frac{\left(\frac{K}{M}-1\right)}{K(2 N-1)-1}}}.
\end{equation}
In general, $\rho\geq 1$. When $\rho=1$, a QCSS is said to be optimal. A QCSS is near-optimal when $1<\rho\leq2$.

\section{Florentine Rectangles and Vatican Squares. Its Existence and Constructions}
In this section we will revisit the definition and existence of Florentine rectangles and Vatican squares. We will also go through some of the constructions of Florentine rectangles and Vatican squares.

\begin{definition}\label{tuscandef}
	A Tuscan-$k$ rectangle of order $r \times N$ has $r$ rows and $N$ columns such that
	\begin{enumerate}
		\item[C1:] each row is a permutation of the $N$ symbols and
		\item[C2:] for any two distinct symbols $a$ and $b$ and for each $1\leq m \leq k$, there is at most one row in which $b$ is $m$ steps to the right of $a$.
	\end{enumerate}
When $k=N-1$, it is called Tuscan-$(N-1)$ rectangle or Florentine rectangle. When $r=N$ and $k=N-1$, it is called Tuscan-$(N-1)$ squares or Florentine squares. 
\end{definition}

\begin{definition}[Latin Square]
	A matrix $\mathcal{A}$ of size $N\times N$ is called a Latin square if each row and each column of $\mathcal{A}$ contains the $N$ symbols, say $0,1,\dots,N-1$, exactly once.
\end{definition}

\begin{definition}
	An $N\times N$ Florentine square is called Vatican square if it is also Latin.
\end{definition}

\begin{example}\label{ex1}
	Examples of $4 \times 4$ Vatican square and $6\times 7$ Florentine rectangle are given below in the form of matrices, respectively.
		\begin{equation}
		\begin{bmatrix}
		0,~1,~3,~2 \\
		1,~2,~0,~3 \\
		2,~3,~1,~0 \\
		3,~0,~2,~1 \\
		\end{bmatrix},
		\begin{bmatrix}
		0,~ 1,~ 2,~ 3,~ 4,~ 5,~ 6 \\
		0,~ 2,~ 4,~ 6,~ 1,~ 3,~ 5 \\
		0,~ 3,~ 6,~ 2,~ 5,~ 1,~ 4 \\
		0,~ 4,~ 1,~ 5,~ 2,~ 6,~ 3 \\
		0,~ 5,~ 3,~ 1,~ 6,~ 4,~ 2 \\
		0,~ 6,~ 5,~ 4,~ 3,~ 2,~ 1 \\
		\end{bmatrix}.
	\end{equation}
	As we see, C1 and C2 of \textit{Definition} \ref{tuscandef} hold here.
\end{example}

For each positive integer $N$ let $F(N)$ denote the largest integer such that Florentine rectangle of order $F(N) \times N$ exists.

There are very few constructions of Florentine rectangles reported in literature. All the reported constructions of the Florentine rectangles are mostly based on extensive computer search. Much of the research reported in the literature is on finding the existence of an $F(N)\times N$ Florentine rectangle for a given $N$. Based on that, for a given value of $0<N\leq 32$, we have given the probable value of $F(N)$, in Table \ref{theoryfn}, for which an $F(N)\times N$ Florentine rectangle can exist.

\begin{table}
	\small
	%	\renewcommand{\arraystretch}{1.3}
%	\resizebox{\textwidth}{!}{
		\caption{Possible values of $F(N)$ for $0<N\leq 32$\label{theoryfn} \cite{song}.}
		\begin{tabular}{|c|c|c|c|}
			\hline
			$N$ & \makecell{Possible value \\of $F(N)$} & $N$ & \makecell{Possible value \\of $F(N)$}  \\
			\hline
			1 & 1 & 17 & 16, 17 \\
			\hline
			2 & 2 & 18 & 18 \\
			\hline
			3 & 2 & 19 & 18, 19 \\
			\hline
			4 & 4 & 20 & 6,$\dots$, 20 \\
			\hline
			5 & 4 & 21 & 6,$\dots$, 21 \\
			\hline
			6 & 6 & 22 & 22 \\
			\hline
			7 & 6 & 23 & 22, 23 \\
			\hline
			8 & 7 & 24 & 6,$\dots$, 24 \\
			\hline
			9 & 8 & 25 & 6,$\dots$, 25 \\
			\hline
			10 & 10 & 26 & 6,$\dots$, 26 \\
			\hline
			11 & 10 & 27 & 6,$\dots$, 27 \\
			\hline
			12 & 12 & 28 & 28 \\
			\hline
			13 & 12,13 & 29 & 28, 29 \\
			\hline
			14 & 6,$\dots$, 14 & 30 & 30 \\
			\hline
			15 & $6,\dots, 15$ & 31 & $30, 31$ \\
			\hline
			16 & 16 & 32 & $6,\dots, 32$ \\
			\hline
\end{tabular} 
\end{table}

\begin{lemma}[Construction of Vatican squares]\label{lem3}
	Let $p$ be an odd prime integer. Then the multiplication table of $\mathbb{Z}_p$, without the border consisting of all-zero row and column, is a $(p-1)\times (p-1)$ Vatican square. This also implies that for $N=p-1$, where $p$ is an odd prime, $F(N)=p-1$.
\end{lemma}

\begin{theorem}\label{thflorentine}
	Let $\mathcal{M}$ be a multiplication table of the elements of $\mathbb{Z}_N$ with the first row $M_{0,j}=0$ for all $0\leq j <N$ and the first column $M_{i,0}=0$ for all $0\leq i<N$. $M_{1,j}=j$ for all $0\leq j <N$ and $M_{i,1}=i$ for all $0\leq i<N$. Let $0<i<p$, $\mathcal{A}=[M_{i,.}]$, where $p$ is the smallest prime factor of $N$. Then $\mathcal{A}$ is a $(p-1)\times N$ Florentine rectangle.
\end{theorem}
\begin{IEEEproof}
	Let the element $b$ be $m$ steps right to $a$ in two rows $r_1$ and $r_2$ of the matrix $\mathcal{A}$, for some $r_1\neq r_2$, $a\neq b$. Let us assume that $a$ is in the $A_{r_1,c_1}$ and $A_{r_2,c_2}$ positions, for some columns $c_1$ and $c_2$ of $\mathcal{A}$. Therefore, according to our assumptions $b$ will be in positions $A_{r_1,c_1+m}$ and $A_{r_2,c_2+m}$, respectively. Since $\mathcal{A}$ is a multiplication table of the residue of $\mathbb{Z}_N$, so we have the following:
	\begin{equation}\label{eq2n}
	a\equiv r_1\cdot c_1 \equiv r_2\cdot c_2 \pmod N.
	\end{equation}
	Similarly, we also have 
	\begin{equation}\label{eq3n}
	b\equiv r_1\cdot (c_1+m) \equiv r_2\cdot (c_2+m) \pmod N.
	\end{equation}
	As per our construction $A_{r_1,1}=r_1$ and $A_{r_2,1}=r_2$. Since $r_1$ and $r_2$ are non-zero residues and prime to $N$ therefore their inverses exist in $\mathbb{Z}_N$. From (\ref{eq2n}) we have 
	\begin{equation}
	\begin{split}
	c_1&\equiv r_1^{-1}\cdot a \pmod N ,\\
	\text{and } c_2&\equiv r_2^{-1}\cdot a \pmod N.
	\end{split}
	\end{equation}
	Replacing the value of $c_1$ and $c_2$ in (\ref{eq3n}) we have 
	\begin{equation}\label{eq5n}
	\begin{split}
	b&\equiv r_1 \cdot (r_1^{-1}\cdot a+m) \pmod N,\\
	\text{also, } b&\equiv r_2 \cdot (r_2^{-1}\cdot a+m) \pmod N.
	\end{split}
	\end{equation}
	Since $m\not \equiv 0\pmod N$, therefore from (\ref{eq5n}) we have,
	\begin{equation}\label{eq6n}
	\begin{split}
	r_1m&\equiv r_2m \pmod N\\
	\implies r_1 &\equiv r_2 \pmod N.
	\end{split}
	\end{equation}
	Since $r_1$ and $r_2$ are less than $p$, the smallest prime factor of $N$, therefore (\ref{eq6n}) implies $r_1=r_2$, or in other words, $r_1$ and $r_2$ are same row, which contradicts our assumption. Hence $\mathcal{A}$ is a $(p-1)\times N$ Florentine rectangle.
\end{IEEEproof}

\begin{theorem}\label{thmm2}
	Let $N\geq 4$, $N=2m$ and $m\not \equiv 1 \pmod 3$. Let $\mathcal{A}$ be a $4 \times N$ matrix and $A_{i,j}$ denotes the $j$-th element of the $i$-th row. Define $A_{i,j}$ as follows:
	\begin{equation}
		A_{0,j}=\begin{cases}
			j \pmod{N+1},~\text{ for }1\leq j \leq m;\\
			(N+1)-A_{0,N+1-j} \pmod{N+1} \\ \hspace{4cm}\text{ for }m+1\leq j \leq N;
		\end{cases}
	\end{equation}
\begin{equation}
	A_{1,j}=\begin{cases}
		2j \pmod{N+1},~\text{ for }1\leq j \leq m;\\
		(N+1)-A_{1,N+1-j} \pmod{N+1} \\ \hspace{4cm}\text{ for }m+1\leq j \leq N;
	\end{cases}
\end{equation}
for $i=2$ and $i=3$, 
\begin{equation}
	A_{i,j}=
		A_{3-i,N+1-j},~\text{ for }1\leq j \leq N.
\end{equation}
	Then $\mathcal{A}$ is a $4\times N$ Florentine rectangle.
\end{theorem}

\begin{IEEEproof}
	Let the element $b$ be $k$ steps right to $a$ in two rows $r_1$ and $r_2$ of the matrix $\mathcal{A}$, for some $r_1\neq r_2$, $a\neq b$. Let us assume that $a$ is in the $A_{r_1,c_1}$ and $A_{r_2,c_2}$ positions, for some columns $c_1$ and $c_2$ of $\mathcal{A}$. Therefore, according to our assumptions $b$ will be in positions $A_{r_1,c_1+k}$ and $A_{r_2,c_2+k}$, respectively. 
	
	Let us consider $r_1=0$ and $r_2=1$, i.e., consider the positions of $a$ and $b$ in the first and second row. When $r_1=0$, if $1\leq c_1 \leq m$ then $a=c_1\pmod{N+1}$, if $m<c_1\leq N$ then $a=(N+1)-(N+1)+c_1\pmod{N+1}=c_1\pmod{N+1}$. According to our assumption in row $r_1=0$, $b$ will be in position $A_{0,c_1+k}$. For $r_1=0$, when $1< c_1+k\leq m$ then $b=c_1+k\pmod{N+1}$, when $m< c_1+k\leq N$ then $b=(N+1)-(N+1)+(c_1+k)\pmod{N+1}=c_1+k\pmod{N+1}$. 
	
	When $r_2=1$, if $1\leq c_2 \leq m$ then $a=2c_2\pmod{N+1}$, if $m<c_2\leq N$ then $a=(N+1)-2(N+1)+2c_2\pmod{N+1}=2c_2\pmod{N+1}$. According to our assumption in row $r_2=1$, $b$ will be in position $A_{1,c_2+k}$. For $r_2=1$, when $1< c_2+k\leq m$ then $b=2(c_2+k)\pmod{N+1}$, when $m< c_2+k\leq N$ then $b=(N+1)-2(N+1)+2(c_1+k)\pmod{N+1}=2(c_1+k)\pmod{N+1}$. 
	
	Following the above detail explanation there will be sixteen sub-cases. The first sub-case is when $c_1\leq m$, $c_1+k\leq m$, $c_2\leq m$, and $c_2+k\leq m$. In this case, we have
	\begin{equation}\label{eq17}
		a\equiv c_1 \equiv 2c_2 \pmod{N+1},
	\end{equation}
and
\begin{equation}\label{eq18}
	b\equiv c_1+k \equiv 2(c_2+k) \pmod{N+1}.
\end{equation}
Combining (\ref{eq17}) and (\ref{eq18}), we get
\begin{equation}
	k\equiv 2k \pmod{N+1},
\end{equation}
which is impossible, since $N\geq 4$ and $k\geq 1$.
Similarly we can prove the other sub-cases. Also similar proof can be given for other values of $r_1$ and $r_2$. Hence $\mathcal{A}$ is a $4\times N$ Florentine rectangle.
\end{IEEEproof}

\begin{theorem}\label{thmm3}
	Let $N\geq 4$, $N=2m$ and $m\not \equiv 0 \pmod 3$. Let $\mathcal{A}$ be a $4 \times N$ matrix and $A_{i,j}$ denotes the $j$-th element of the $i$-th row. Define $A_{i,j}$ as follows:
	\begin{equation}
		A_{0,j}=\begin{cases}
			j,&\text{ for }1\leq j \leq m;\\
			(N+1)-A_{0,N+1-j} &\text{ for }m+1\leq j \leq N;
		\end{cases}
	\end{equation}
	\begin{equation}
		A_{1,j}=\begin{cases}
			1+\lbrace[2(i-1)+m-1]\pmod N \rbrace,\\ \hspace{4cm} \text{ for }1\leq j \leq m;\\
			(N+1)-A_{1,N+1-j}, \\ \hspace{4cm} \text{ for }m+1\leq j \leq N;
		\end{cases}
	\end{equation}
	for $i=2$ and $i=3$, 
	\begin{equation}
		A_{i,j}=
		A_{3-i,N+1-j},~\text{ for }1\leq j \leq N.
	\end{equation}
	Then $\mathcal{A}$ is a $4\times N$ Florentine rectangle.
\end{theorem}

\begin{IEEEproof}
	Let the element $b$ be $k$ steps right to $a$ in two rows $r_1$ and $r_2$ of the matrix $\mathcal{A}$, for some $r_1\neq r_2$, $a\neq b$. Let us assume that $a$ is in the $A_{r_1,c_1}$ and $A_{r_2,c_2}$ positions, for some columns $c_1$ and $c_2$ of $\mathcal{A}$. Therefore, according to our assumptions $b$ will be in positions $A_{r_1,c_1+k}$ and $A_{r_2,c_2+k}$, respectively. 
	
	Let us consider $r_1=0$ and $r_2=1$, i.e., consider the positions of $a$ and $b$ in the first and second row. When $r_1=0$, if $1\leq c_1 \leq m$ then $a=c_1$, if $m<c_1\leq N$ then $a=(N+1)-(N+1)+c_1=c_1$. According to our assumption in row $r_1=0$, $b$ will be in position $A_{0,c_1+k}$. For $r_1=0$, when $1< c_1+k\leq m$ then $b=c_1+k$, when $m< c_1+k\leq N$ then $b=(N+1)-(N+1)+(c_1+k)=c_1+k$. 
	
	When $r_2=1$, if $1\leq c_2 \leq m$ then $a=1+\{[2(c_2-1)+m-1]\pmod N\}$, if $m<c_2\leq N$ then $a=(N+1)-1-\{[2(N+1-c_2-1)+m-1]\pmod N\}=N-\{[2(N-c_2)+m-1]\pmod N\}$. According to our assumption in row $r_2=1$, $b$ will be in position $A_{1,c_2+k}$. For $r_2=1$, when $1< c_2+k\leq m$ then $b=1+\{[2(c_2+k-1)+m-1]\pmod N\}$, when $m< c_2+k\leq N$ then $b=(N+1)-1-\{[2(N+1-c_2-k-1)+m-1]\pmod N\}=N-\{[2(N-c_2-k)+m-1]\pmod N\}$.
	
	Following the above detail explanation there will be sixteen sub-cases. The first sub-case is when $c_1\leq m$, $c_1+k\leq m$, $c_2\leq m$, and $c_2+k\leq m$. In this case, we have
	\begin{equation}\label{eq171}
		a\equiv c_1 \equiv 1+\{[2(c_2-1)+m-1]\pmod N\},
	\end{equation}
	and
	\begin{equation}\label{eq181}
		\begin{split}
					b&\equiv c_1+k \equiv 1+\{[2(c_2+k-1)+m-1]\pmod N\}\\
					&\implies c_1+k-1\equiv \{\{[2(c_2-1)+m-1]\pmod N \\& \hspace{3cm}+2k \pmod N\}\pmod N\}
		\end{split}
	\end{equation}
From (\ref{eq171}) we can write (\ref{eq181}) as 
\begin{equation}\label{eq1812}
	c_1+k-1\equiv \{[(c_1-1)+2k\pmod N] \pmod N\}.
\end{equation}
Since $c_1+k\leq m$, therefore $2k \pmod N=2k$, and hence (\ref{eq1812}) can be written as
\begin{equation}
	\begin{split}
		&c_1+k-1\equiv \{[(c_1-1)+2k] \pmod N\}\\
		&\implies c_1-1+2k=NQ+c_1+k-1,
	\end{split}
\end{equation}
For some integer $Q$, and hence
\begin{equation}
	k=NQ,
\end{equation}
which is impossible, since $k<N$. Hence our assumption is impossible. Similarly we can prove the other sub-cases. Also similar proof can be given for other values of $r_1$ and $r_2$. Hence $\mathcal{A}$ is a $4\times N$ Florentine rectangle.
\end{IEEEproof}

\begin{corollary}\label{corrr1}
	Let $N=2m+1$. Using \textit{Theorem} \ref{thmm2} and \textit{Theorem} \ref{thmm3}, we can construct a $4\times N$ Florentine rectangle for this case also, just by adding a column of all $N$'s at the end of the rectangle.
\end{corollary}

\begin{remark}
	For a given $N$, the values given in Table \ref{theoryfn} are higher. However, systematic construction of Florentine rectangles of order $F(N)\times N$, when $F(N)$ achieves the maximum value, is only available when $N=p$ or $N=p-1$, where $p$ is prime. Computer search results of Florentine rectangles of order $F(N)\times N$, are available for small values of $N$. Although \textit{Theorem \ref{thflorentine}}, \textit{Theorem \ref{thmm2}}, \textit{Theorem \ref{thmm3}} or \textit{Corollary \ref{corrr1}} do not always results to maximum number of rows for a given $N$ which one can obtain by a computer search, we will use these results to construct the Florentine rectangles, since these are systematic constructions.
\end{remark}

\begin{remark}\label{rem2}
	For a given $N$, in this paper, we consider the value of $F(N)$ as the largest integer for which we can systematically construct a Florentine rectangle of size $F(N)\times N$. The Florentine rectangles are constructed as per following:
	\begin{enumerate}
		\item For $N=p-1$, where $p$ is prime, we construct the permutations as per \textit{Lemma} \ref{lem3}, and then subtracting $1$ from each of the elements. Therefore $F(p-1)=p-1$.
		\item For $N=p$, where $p$ is prime, we first construct the Vatican square as per \textit{Lemma} \ref{lem3} and then add zero column in the left side of the Vatican square. Therefore $F(p)=p-1$.
		\item Let $N>5$ be any number where $N=p_0^{r_0}p_1^{r_1}\dots p_{n-1}^{r_{n-1}}$ and $p_0$ be the least prime factor of $N$. We also consider $N+1$, where 
		$N+1=e_0^{s_0}e_1^{s_1}\dots e_{m-1}^{s_{m-1}}$ and $e_0$ is the least prime factor of $N+1$. Assume $p_0\neq 2,3$ and $e_0\neq 2,3$. If $p_0> e_0$, we follow \textit{Theorem} \ref{thflorentine} to construct the Florentine rectangle and eventually the permutations over $N$. Therefore $F(N)=p_0-1$. If $p_0< e_0$, we follow \textit{Theorem} \ref{thflorentine} to construct the Florentine rectangle and eventually the permutations over $N+1$. Then we remove the all zero column from the rectangle and subtract $1$ from each element. Therefore, $F(N)=e_0-1$.
		\item For any positive integer $N$ having the least prime factor $2$ or $3$ and does not fall under above three cases, we follow \textit{Theorem \ref{thmm2}}, \textit{Theorem \ref{thmm3}} or \textit{Corollary \ref{corrr1}}. Therefore, $F(N)=4$.
	\end{enumerate} 
The value of $F(N)$ considered in this paper is strictly based on the above set of rules.
\end{remark}

\section{Multiple CCCs From Florentine Rectangles}
% Inspired by the construction given in
In this section, first we analyse some intrinsic properties of the Florentine rectangles and then we will utilise those properties to construct several sets of CCCs with low inter-set cross-correlation magnitude. Let us begin by recalling the following property of the Florentine rectangles.

%The main idea behind the construction is inspired by the recent works in \cite{Li19_4}, \cite{zhou18} and \cite{zhou2020}.
%%%%%%%%%%%%%%%%%%%%%%%%%%%%%%%%%%%%%%%%%%%%%%%%%%%%%%%%
\begin{property}\label{prop1}
	Let $\mathcal{A}$ be a $F(N)\times N$ Florentine rectangle over $\mathbb{Z}_N$, denoted as follows
	\begin{equation}
		\mathcal{A}=\left[ \begin{array}{c}{A_{0,0},\dots,A_{0,N-1} } \\ {A_{1,0},\dots,A_{1,N-1} } \\ {\vdots} \\ {A_{F(N)-1,0},\dots,A_{F(N)-1,N-1} }\end{array}\right]_{F(N) \times N}, 
	\end{equation}
	where $A_{i,j}$ denotes the $j$-th element in the $i$-th row. According to Definition \ref{tuscandef}, each row of $\mathcal{A}$, i.e., $A_i$, is a permutation on $\mathbb{Z}_N$. Also, for each $0<\alpha<N$, $(A_{i,j},A_{i,j+\alpha})\neq (A_{m,n},A_{m,n+\alpha })$ unless $i=m$ and $j=n$, where $0\leq i,m \leq F(N)-1$, $0\leq j,n \leq N-1$, $0<j+\alpha<N$ and $0<n+\alpha<N$. In other words, if $\pi_i:\mathbb{Z}_N\rightarrow \mathbb{Z}_N$ be a permutation on $\mathbb{Z}_N$, where $\pi_i$ is equivalent to the $i$-th row of $\mathcal{A}$, i.e. $A_i$, then for each $0<\alpha<N$ $(\pi_i(j),\pi_i(j+\alpha))=(\pi_m(n),\pi_m(n+\alpha ))$ if and only if $i=m$ and $j=n$, where $0<j+\alpha<N$ and $0<n+\alpha<N$.
\end{property}

\begin{lemma}\label{lem_new}
	Let $\pi_i:\mathbb{Z}_N\rightarrow \mathbb{Z}_N$ be a permutation over $\mathbb{Z}_N$ as defined in Property \ref{prop1}. For $0\leq i\neq m \leq F(N)-1$, $\pi_i(j)=\pi_m(j+\tau)$, where $0\leq j+\tau<N$, has at most one solution for $0\leq \tau<N$.
\end{lemma}
\begin{IEEEproof}
	Let us assume that for $0\leq i\neq m \leq F(N)-1$, $\pi_i(j)=\pi_m(j+\tau)$, where $0\leq j+\tau<N$, has more than one solution for $0\leq \tau<N$. Let two of the solutions be $j_1$ and $j_2$. Then $\pi_i(j_1)=\pi_m(j_1+\tau )$ and $\pi_i(j_2)=\pi_m(j_2+\tau )$. Therefore, we have $(\pi_i(j_1),\pi_i(j_2))=(\pi_m(j_1+\tau ),\pi_m(j_2+\tau ))$. This contradicts the definition of Florentine rectangles. Hence, $\pi_i(j)=\pi_m(j+\tau)$ has at most one solution for each $0<\tau<N$ for $0\leq i\neq m \leq F(N)-1$. 
\end{IEEEproof}

 The following example illustrates the set of permutations on $\mathbb{Z}_{N}$ defined by the Florentine rectangles.
\begin{example}
	Consider $N=4$. From Example \ref{ex1}, we have
	$\pi_0(\mathbb{Z}_4)=\{0,1,3,2\}$, $\pi_1(\mathbb{Z}_4)=\{1,2,0,3\}$, $\pi_2(\mathbb{Z}_4)=\{2,3,1,0\}$, $\pi_3(\mathbb{Z}_4)=\{3,0,2,1\}$.
\end{example}

Utilizing the set of permutations defined in Property \ref{prop1}, we design several sets of CCCs using the construction framework given below.

\begin{construction}\label{new_const}
		Consider any positive integer $N\geq 2$, for which an $F(N)\times N$ Florentine rectangle $\mathcal{A}$ exists over $\mathbb{Z}_N$. Also let $\pi_k$ be the permutation over $\mathbb{Z}_N$ for $0\leq k <F(N)$, defined as above, which satisfies Lemma \ref{lem_new}. Then for each $m\in  \mathbb{Z}_N$, and $s\in \mathbb{Z}_N$, define $f_{s}^{(k,m)}:\mathbb{Z}_N\rightarrow \mathbb{Z}_N$ as follows
	\begin{equation}\label{eq_new_const2}
	f_{s}^{(k,m)}(t)=
	s\cdot \pi_k(t)+m t \pmod {N}.
	\end{equation}
		For each $0 \leq k <F(N)$, define a set
	\begin{equation}\label{C^k_2}
	\mathcal{\mathfrak{C}}^{k}=\{\mathcal{C}^{(k,0)},\mathcal{C}^{(k,1)},\cdots,
	\mathcal{C}^{(k,N-1)}\},
	\end{equation}
	where
	\[
	\mathcal{C}^{(k,m)}\hspace{-0.1cm}=\hspace{-0.1cm}
	\begin{bmatrix}
	C_{0}^{(k,m)}\\
	C_{1}^{(k,m)}\\
	\vdots  \\
	C_{N-1}^{(k,m)} \\
	\end{bmatrix}\hspace{-0.1cm}
	=\hspace{-0.1cm}
	\begin{bmatrix}
	C_{0,0}^{(k,m)} , C_{0,1}^{(k,m)} , \cdots , C_{0,N-1}^{(k,m)} \\
	C_{1,0}^{(k,m)} , C_{1,1}^{(k,m)} , \cdots , C_{1,N-1}^{(k,m)} \\
	\vdots  \\
	C_{N-1,0}^{(k,m)} , C_{N-1,1}^{(k,m)} , \cdots , C_{N-1,N-1}^{(k,m)} \\
	\end{bmatrix}
	\]
	and
	$$
	C_{s,t}^{(k,m)}=\omega_{N}^{f_{s}^{(k,m)}(t)} \textrm{~for~each~}0\leq t\leq N-1.
	$$
\end{construction}

For the sequence sets obtained using Construction \ref{new_const}, we have the following theorem.
\begin{theorem}\label{theorem2}
	For $0 \leq k <F(N)$, let $\mathfrak{C}^{k}$ be the multiple sequence sets obtained from Construction \ref{new_const} based on the function $f_{s}^{(k,m)}: \mathbb{Z}_{N}\rightarrow\mathbb{Z}_{N}$, as given in (\ref{eq_new_const2}). Then,
	\begin{enumerate}
		\item For each $0\leq k <F(N)$, $\mathfrak{C}^{k}$ is an $(N,N,N)$- CCC.
		\item For any two different CCCs ${\mathfrak{C}}^{k_1}$ and ${\mathfrak{C}}^{k_2}$, the inter-set cross-correlation is upper bounded by $N$, i.e.,
		\begin{equation}
			|\sum\limits_{s=0}^{N-1} \tilde{R}_{C_{s}^{\left(k_{1}, m_{1}\right)},C_{s}^{\left(k_{2}, m_{2}\right)} }(\tau)| \leq N,
		\end{equation} 
		for all $0\leq k_1\neq k_2 <F(N)$, $0\leq\tau \leq N-1$, and $0\leq m_1,m_2\leq N-1.$
	\end{enumerate}
\end{theorem}

\begin{IEEEproof}
	The proof is given in Appendix A.
\end{IEEEproof}

\begin{example}\label{exn9new}
	Let $N=10$, The family of permutations being defined as in the first rule of \textit{Remark \ref{rem2}}. Then we have
	\begin{equation}
	\begin{split}
	\pi_0(\mathbb{Z}_{10})&=\{0,~1,~2,~3,~4,~5,~6,~7,~8,~9\},\\
	\pi_1(\mathbb{Z}_{10})&=\{1,~3,~5,~7,~9,~0,~2,~4,~6,~8\}, \\
	\pi_2(\mathbb{Z}_{10})&=\{2,~5,~8,~0,~3,~6,~9,~1,~4,~7\}, \\
	\pi_3(\mathbb{Z}_{10})&=\{3,~7,~0,~4,~8,~1,~5,~9,~2,~6\},\\
	\pi_4(\mathbb{Z}_{10})&=\{4,~9,~3,~8,~2,~7,~1,~6,~0,~5\},\\
	\pi_5(\mathbb{Z}_{10})&=\{5,~0,~6,~1,~7,~2,~8,~3,~9,~4\},\\
	\pi_6(\mathbb{Z}_{10})&=\{6,~2,~9,~5,~1,~8,~4,~0,~7,~3\},\\
	\pi_7(\mathbb{Z}_{10})&=\{7,~4,~1,~9,~6,~3,~0,~8,~5,~2\}, \\
	\pi_8(\mathbb{Z}_{10})&=\{8,~6,~4,~2,~0,~9,~7,~5,~3,~1\}, \\
	\pi_9(\mathbb{Z}_{10})&=\{9,~8,~7,~6,~5,~4,~3,~2,~1,~0\}.
	\end{split}
	\end{equation}
	By Theorem \ref{theorem2}, we get ten $(10,10,10)$-CCCs, $\mathfrak{C}^{0}, \mathfrak{C}^{1}, \mathfrak{C}^{2}, \mathfrak{C}^{3}, \mathfrak{C}^{4},\mathfrak{C}^{5},\mathfrak{C}^{6},\mathfrak{C}^{7},\mathfrak{C}^{8},\mathfrak{C}^{9}$. The two CCCs $\mathfrak{C}^{2}$ and $\mathfrak{C}^{3}$ are shown in Table \ref{tab_ex2}.
\end{example}

\begin{table*}[]
		\small
	\resizebox{\textwidth}{!}{
	\begin{tabular}{|c|c|c|c|c|c|c|c|c|c|}
		\hline
		$\mathcal{C}^{(2,0)}$                                                                                                                                                   & $\mathcal{C}^{(2,1)}$                                                                                                                                                   & $\mathcal{C}^{(2,2)}$                                                                                                                                                   & $\mathcal{C}^{(2,3)}$                                                                                                                                                   & $\mathcal{C}^{(2,4)}$                                                                                                                                                   & $\mathcal{C}^{(2,5)}$                                                                                                                                                   & $\mathcal{C}^{(2,6)}$                                                                                                                                                   & $\mathcal{C}^{(2,7)}$                                                                                                                                                   & $\mathcal{C}^{(2,8)}$                                                                                                                                                   & $\mathcal{C}^{(2,9)}$                                                                                                                                                   \\ \hline
		\begin{tabular}[c]{@{}c@{}}0000000000\\ 2580369147\\ 4060628284\\ 6540987321\\ 8020246468\\ 0500505505\\ 2080864642\\ 4560123789\\ 6040482826\\ 8520741963\end{tabular} & \begin{tabular}[c]{@{}c@{}}0123456789\\ 2603715826\\ 4183074963\\ 6663333000\\ 8143692147\\ 0623951284\\ 2103210321\\ 4683579468\\ 6163838505\\ 8643197642\end{tabular} & \begin{tabular}[c]{@{}c@{}}0246802468\\ 2726161505\\ 4206420642\\ 6786789789\\ 8266048826\\ 0746307963\\ 2226666000\\ 4706925147\\ 6286284284\\ 8766543321\end{tabular} & \begin{tabular}[c]{@{}c@{}}0369258147\\ 2849517284\\ 4329876321\\ 6809135468\\ 8389494505\\ 0869753642\\ 2349012789\\ 4829371826\\ 6309630963\\ 8889999000\end{tabular} & \begin{tabular}[c]{@{}c@{}}0482604826\\ 2962963963\\ 4442222000\\ 6922581147\\ 8402840284\\ 0982109321\\ 2462468468\\ 4942727505\\ 6422086642\\ 8902345789\end{tabular} & \begin{tabular}[c]{@{}c@{}}0505050505\\ 2085319642\\ 4565678789\\ 6045937826\\ 8525296963\\ 0005555000\\ 2585814147\\ 4065173284\\ 6545432321\\ 8025791468\end{tabular} & \begin{tabular}[c]{@{}c@{}}0628406284\\ 2108765321\\ 4688024468\\ 6168383505\\ 8648642642\\ 0128901789\\ 2608260826\\ 4188529963\\ 6668888000\\ 8148147147\end{tabular} & \begin{tabular}[c]{@{}c@{}}0741852963\\ 2221111000\\ 4701470147\\ 6281739284\\ 8761098321\\ 0241357468\\ 2721616505\\ 4201975642\\ 6781234789\\ 8261593826\end{tabular} & \begin{tabular}[c]{@{}c@{}}0864208642\\ 2344567789\\ 4824826826\\ 6304185963\\ 8884444000\\ 0364703147\\ 2844062284\\ 4324321321\\ 6804680468\\ 8384949505\end{tabular} & \begin{tabular}[c]{@{}c@{}}0987654321\\ 2467913468\\ 4947272505\\ 6427531642\\ 8907890789\\ 0487159826\\ 2967418963\\ 4447777000\\ 6927036147\\ 8407395284\end{tabular} \\ \hline
		$\mathcal{C}^{(3,0)}$                                                                                                                                                   & $\mathcal{C}^{(3,1)}$                                                                                                                                                   & $\mathcal{C}^{(3,2)}$                                                                                                                                                   & $\mathcal{C}^{(3,3)}$                                                                                                                                                   & $\mathcal{C}^{(3,4)}$                                                                                                                                                   & $\mathcal{C}^{(3,5)}$                                                                                                                                                   & $\mathcal{C}^{(3,6)}$                                                                                                                                                   & $\mathcal{C}^{(3,7)}$                                                                                                                                                   & $\mathcal{C}^{(3,8)}$                                                                                                                                                   & $\mathcal{C}^{(3,9)}$                                                                                                                                                   \\ \hline
		\begin{tabular}[c]{@{}c@{}}0000000000\\ 3704815926\\ 6408620842\\ 9102435768\\ 2806240684\\ 5500055500\\ 8204860426\\ 1908675342\\ 4602480268\\ 7306295184\end{tabular} & \begin{tabular}[c]{@{}c@{}}0123456789\\ 3827261605\\ 6521076521\\ 9225881447\\ 2929696363\\ 5623401289\\ 8327216105\\ 1021021021\\ 4725836947\\ 7429641863\end{tabular} & \begin{tabular}[c]{@{}c@{}}0246802468\\ 3940617384\\ 6644422200\\ 9348237126\\ 2042042042\\ 5746857968\\ 8440662884\\ 1144477700\\ 4848282626\\ 7542097542\end{tabular} & \begin{tabular}[c]{@{}c@{}}0369258147\\ 3063063063\\ 6767878989\\ 9461683805\\ 2165498721\\ 5869203647\\ 8563018563\\ 1267823489\\ 4961638305\\ 7665443221\end{tabular} & \begin{tabular}[c]{@{}c@{}}0482604826\\ 3186419742\\ 6880224668\\ 9584039584\\ 2288844400\\ 5982659326\\ 8686464242\\ 1380279168\\ 4084084084\\ 7788899900\end{tabular} & \begin{tabular}[c]{@{}c@{}}0505050505\\ 3209865421\\ 6903670347\\ 9607485263\\ 2301290189\\ 5005005005\\ 8709810921\\ 1403625847\\ 4107430763\\ 7801245689\end{tabular} & \begin{tabular}[c]{@{}c@{}}0628406284\\ 3322211100\\ 6026026026\\ 9720831942\\ 2424646868\\ 5128451784\\ 8822266600\\ 1526071526\\ 4220886442\\ 7924691368\end{tabular} & \begin{tabular}[c]{@{}c@{}}0741852963\\ 3445667889\\ 6149472705\\ 9843287621\\ 2547092547\\ 5241807463\\ 8945612389\\ 1649427205\\ 4343232121\\ 7047047047\end{tabular} & \begin{tabular}[c]{@{}c@{}}0864208642\\ 3568013568\\ 6262828484\\ 9966633300\\ 2660448226\\ 5364253142\\ 8068068068\\ 1762873984\\ 4466688800\\ 7160493726\end{tabular} & \begin{tabular}[c]{@{}c@{}}0987654321\\ 3681469247\\ 6385274163\\ 9089089089\\ 2783894905\\ 5487609821\\ 8181414747\\ 1885229663\\ 4589034589\\ 7283849405\end{tabular} \\ \hline
	\end{tabular}}
\caption{The two $(10,10,10)$- CCCs $\mathfrak{C}^2$ and $\mathfrak{C}^3$ of \textit{Example \ref{exn9new}}.\label{tab_ex2}}
\end{table*}

%\begin{example}\label{exn9}
%	Let $N=9$, The family of permutations being defined as in Example \ref{ex1}. Then we have
%	\begin{equation}
%		\begin{split}
%		\pi_0(\mathbb{Z}_9)&=\{0,~ 1,~ 2,~ 3,~ 4,~ 5,~ 6,~ 7,~ 8\},\\
%		\pi_1(\mathbb{Z}_9)&=\{1,~ 8,~ 5,~ 0,~ 4,~ 7,~ 2,~ 6,~ 3\}, \\
%		\pi_2(\mathbb{Z}_9)&=\{2,~ 5,~ 8,~ 6,~ 4,~ 3,~ 1,~ 0,~ 7\}, \\
%		\pi_3(\mathbb{Z}_9)&=\{3,~ 0,~ 6,~ 8,~ 4,~ 2,~ 7,~ 5,~ 1\},\\
%		\pi_4(\mathbb{Z}_9)&=\{5,~ 7,~ 3,~ 2,~ 4,~ 0,~ 8,~ 1,~ 6\},\\
%		\pi_5(\mathbb{Z}_9)&=\{6,~ 2,~ 1,~ 7,~ 4,~ 8,~ 0,~ 3,~ 5\},\\
%		\pi_6(\mathbb{Z}_9)&=\{7,~ 6,~ 0,~ 5,~ 4,~ 1,~ 3,~ 8,~ 2\},\\
%		\pi_7(\mathbb{Z}_9)&=\{8,~ 3,~ 7,~ 1,~ 4,~ 6,~ 5,~ 2,~ 0\}.
%		\end{split}
%	\end{equation}
%			By Theorem \ref{theorem2}, we get eight $(9,9,9)$-CCCs, $\mathfrak{C}^{1}, \mathfrak{C}^{2}, \mathfrak{C}^{3}, \mathfrak{C}^{4}, \mathfrak{C}^{5},\mathfrak{C}^{6}, \mathfrak{C}^{7},\mathfrak{C}^{8}$. The first complementary set of first two CCCs are shown in Table \ref{tab_ex2}.
%\end{example}

We propose a framework to systematically construct asymptotically optimal aperiodic QCSSs, in the following section.

\section{Proposed Construction of Asymptotically Optimal QCSSs}

\begin{theorem}\label{cor_n_5}
Consider $N$ to be any positive integer greater than $3$, for which $F(N)>4$. Also, let $\mathfrak{C}^{k}\text{ for }0\leq k < F(N)$ be obtained using Theorem \ref{theorem2} and
$\mathfrak{C}=\mathfrak{C}^{0} \cup \mathfrak{C}^{1} \cup \cdots \cup \mathfrak{C}^{F(N)-1}$, then the sequence set $\mathfrak{C}$ is an asymptotically optimal aperiodic $(N\times F(N),N,N,N)$-QCSS.
\end{theorem}

\begin{IEEEproof}
	Based on Theorem \ref{theorem2}, for each $0\leq k< F(N)$, $\mathfrak{C}^{k}$ is an $(N,N,N)$-CCC. Also, for any two $k_1$, $k_2$, where $0\leq k_1,k_2< F(N)$ and $k_1\neq k_2$, the the inter-set cross-correlation magnitude between $\mathfrak{C}^{k_1}$ and $\mathfrak{C}^{k_2}$ is upper bounded by $N$. Then $\mathfrak{C}$ is an aperiodic $(K,M,N,\delta_{\max })$- QCSS, where $K=N\times F(N), ~M=N,~ N=N,~\delta_{\max }=max\{0,N\}=N.$ \par
	
	The optimality factor of $(N\times F(N),N,N,N)$-QCSS is
	
	\begin{equation}\label{neq21}
		 \rho =\frac{N}{\sqrt{N^{2} (1-2\sqrt{\frac{N}{3N\times F(N)}} )}}.
	\end{equation}
When $N\rightarrow +\infty$ then $F(N)\rightarrow +\infty$. Therefore from (\ref{neq21}), 
\begin{equation}
\begin{aligned}\lim_{F(N)\rightarrow +\infty} \rho &=\lim_{F(N)\rightarrow +\infty}\frac{N}{\sqrt{N^{2} (1-2\sqrt{\frac{N}{3N\times F(N)}} )}} \\ &=\lim_{F(N)\rightarrow +\infty}\frac{1}{\sqrt{1-\frac{2}{\sqrt{3\times F(N)}}}}\\&=1.\\
\end{aligned}
\end{equation}
Therefore, $\mathfrak{C}$ is an asymptotically optimal aperiodic QCSS.
\end{IEEEproof}

\begin{example}\label{exm5}
	The ten $(10,10,10)$-CCCs, $\mathfrak{C}^{0}, \mathfrak{C}^{1}, \mathfrak{C}^{2}, \mathfrak{C}^{3}, \mathfrak{C}^{4},\mathfrak{C}^{5},\mathfrak{C}^{6},\mathfrak{C}^{7},\mathfrak{C}^{8},\mathfrak{C}^{9}$, generated in \textit{Example \ref{exn9new}} can be used to construct asymptotically optimal (100,10,10,10)- QCSS $\mathfrak{C}=\mathfrak{C}^{0}\cup\mathfrak{C}^{1}\cup \mathfrak{C}^{2}\cup \mathfrak{C}^{3}\cup \mathfrak{C}^{4}\cup \mathfrak{C}^{5}\cup\mathfrak{C}^{6}\cup\mathfrak{C}^{7}\cup\mathfrak{C}^{8}\cup\mathfrak{C}^{9}$. 
%	A glimpse of the auto-correlation and crosscorrelation of the QCSS is shown in Fig. {\ref{fig1}}.
\end{example}
%\begin{figure}
%	\includegraphics[draft=false,width=\textwidth]{fignten1.eps}
%	\caption{Correlation magnitudes of the sequence sets $\mathcal{C}^{(2,0)}$ and $\mathcal{C}^{(3,0)}$ of the QCSS in \textit{Example \ref{exm5}}.}\label{fig1}
%\end{figure}

The asymptotically optimal aperiodic QCSSs obtained using Theorem \ref{cor_n_5} for some $N$, are shown in Table \ref{tab_composite}, with corresponding parameters.

\begin{table}[h!]
	\small
	%\renewcommand{\arraystretch}{1.3}
	%\resizebox{0.9\columnwidth}{!}{
	\begin{tabular}{|c|c|c|c|c|c|}
	\hline
Alphabet & $K$ & $M$ & $N$ & $\rho$   \\
\hline
$\mathbb{Z}_{6}$ & 36 & 6 & 6 & 1.3754  \\
\hline
$\mathbb{Z}_{10}$ & 100 & 10 & 10 & 1.2551  \\
\hline
$\mathbb{Z}_{12}$ & 144 & 12 & 12 & 1.2247  \\
\hline
$\mathbb{Z}_{18}$ & 324 & 18 & 18 & 1.1722  \\
\hline
$\mathbb{Z}_{22}$ & 484 & 22 & 22 & 1.1518  \\
\hline
$\mathbb{Z}_{28}$ & 784 & 28 & 28 & 1.1310  \\
\hline
$\mathbb{Z}_{30}$ & 900 & 30 & 30 & 1.1257  \\
\hline
$\mathbb{Z}_{36}$ & 1296 & 36 & 36 & 1.1128  \\
\hline
$\mathbb{Z}_{40}$ & 1600 & 40 & 40 & 1.1061  \\
\hline
$\mathbb{Z}_{42}$ & 1764 & 42 & 42 & 1.1031  \\
\hline
$\mathbb{Z}_{46}$ & 2116 & 46 & 46 & 1.0978  \\
\hline
$\mathbb{Z}_{48}$ & 288 & 48 & 48 & 1.3754  \\
\hline
$\mathbb{Z}_{52}$ & 2704 & 52 & 52 & 1.0912  \\
\hline
$\mathbb{Z}_{58}$ & 3364 & 58 & 58 & 1.0857  \\
\hline
$\mathbb{Z}_{60}$ & 3600 & 60 & 60 & 1.0841  \\
\hline
$\mathbb{Z}_{66}$ & 4356 & 66 & 66 & 1.0797  \\
\hline
$\mathbb{Z}_{70}$ & 4900 & 70 & 70 & 1.0771  \\
\hline
$\mathbb{Z}_{72}$ & 5184 & 72 & 72 & 1.0759  \\
\hline
$\mathbb{Z}_{76}$ & 456 & 76 & 76 & 1.3754  \\
\hline
$\mathbb{Z}_{78}$ & 6084 & 78 & 78 & 1.0726  \\
\hline
$\mathbb{Z}_{82}$ & 6724 & 82 & 82 & 1.0706  \\
\hline
$\mathbb{Z}_{88}$ & 7744 & 88 & 88 & 1.0679  \\
\hline
$\mathbb{Z}_{90}$ & 540 & 90 & 90 & 1.3754  \\
\hline
$\mathbb{Z}_{96}$ & 9216 & 96 & 96 & 1.0647  \\
\hline
$\mathbb{Z}_{100}$ & 10000 & 100 & 100 & 1.0633  \\
\hline
	\end{tabular}\\
	\caption{Asymptotically optimal aperiodic QCSSs, when $N$ is even number.\label{tab_composite}}
\end{table}

\begin{table}[h!]
	\small
	%\renewcommand{\arraystretch}{1.3}
	%\resizebox{0.9\columnwidth}{!}{
	\begin{tabular}{|c|c|c|c|c|c|}
		\hline
		Alphabet & $K$ & $M$ & $N$ & $\rho$   \\
		\hline
		$\mathbb{Z}_{14}$ & 56 & 14 & 14 & 1.5382  \\
		\hline
		$\mathbb{Z}_{20}$ & 80 & 20 & 20 & 1.5382  \\
		\hline
		$\mathbb{Z}_{24}$ & 96 & 24 & 24 & 1.5382  \\
		\hline
		$\mathbb{Z}_{26}$ & 104 & 26 & 26 & 1.5382  \\
		\hline
		$\mathbb{Z}_{36}$ & 144 & 36 & 36 & 1.5382  \\
		\hline
		$\mathbb{Z}_{38}$ & 152 & 38 & 38 & 1.5382  \\
		\hline
		$\mathbb{Z}_{44}$ & 176 & 44 & 44 & 1.5382  \\
		\hline
		$\mathbb{Z}_{50}$ & 200 & 50 & 50 & 1.5382  \\
		\hline
	\end{tabular}\\
	\caption{Near-optimal aperiodic QCSSs, when $N$ is even number.\label{tab_composite1}}
\end{table}

\begin{corollary}
	For the cases when $F(N)=4$, $\mathfrak{C}$ is an aperiodic near-optimal $(4N,N,N,N)$-QCSS.
\end{corollary}
\begin{IEEEproof}
From (\ref{neq21}), we get
	\begin{equation}
		\rho =\frac{N}{\sqrt{N^{2} (1-2\sqrt{\frac{N}{3\times N\times 4}} )}}=1.5382 .
	\end{equation}
\end{IEEEproof}

\begin{corollary}\label{case_for_3}
Let $N=2$ or $3$. Also, let $\mathfrak{C}^{0},~\mathfrak{C}^{1}$ be obtained from Theorem \ref{theorem2} and
$\mathfrak{C}=\mathfrak{C}^{0} \cup \mathfrak{C}^{1}$. Then $\mathfrak{C}$ is a near-optimal aperiodic $(4,2,2,2)$-QCSS or $(6,3,3,3)$- QCSS, with optimality factor $\rho=1.6584$ or $1.7950$, respectively. Since $K \ngeq 3M$, we have used the Welch bound, discussed in Lemma \ref{lem1}.	
\end{corollary}

%\begin{IEEEproof}
%	The proof is similar to Theorem \ref{cor_n_5}. However, in this case we will use the Welch bound, given in Lemma \ref{lem1}, since $K \ngeq 3M$.
%	Therefore, the optimality factor of $(2N,N,N,N)$-QCSS is
%	
%	\begin{equation}\label{neq23}
%		\rho =\frac{N}{N^2\sqrt{\frac{(\frac{2N}{N}-1)}{2N(2N-1)-1}}}.
%	\end{equation}
%	Taking limit on both sides of (\ref{neq23}) when $p_0\rightarrow +\infty$, we get
%	\begin{equation}
%	\begin{aligned} \lim_{N\rightarrow +\infty}\rho &=\lim_{N\rightarrow +\infty}\frac{N}{N^2\sqrt{\frac{(\frac{2N}{N}-1)}{2N(2N-1)-1}}} \\ &=\lim_{N\rightarrow +\infty}\frac{1}{\sqrt{\frac{N^2}{4N^2-2N-1}}}\\&=\lim_{N\rightarrow +\infty} \sqrt{4-\frac{2}{N}-\frac{1}{N^2}}\\&=2.\\
%	\end{aligned}
%	\end{equation}
%\end{IEEEproof}

%Based on Corollary \ref{case_for_3}, a list of parameters of asymptotically near-optimal aperiodic QCSSs are given in Table \ref{new_table_for_3}.

\begin{table}[t]
	\small
%	\renewcommand{\arraystretch}{1.3}
%	\resizebox{\textwidth}{!}{
	\begin{tabular}{|c|c|c|c|c|c|c|c|}
		\hline
		Alphabet &$K$& $K_{\_prev}$ & $M$ & $N$ & $\rho$&$\rho_{\_prev}$   \\
		\hline
		$\mathbb{Z}_{3*5}$ &60& 30 & 15 & 15 &1.5382& 1.9653 \\
		\hline
		$\mathbb{Z}_{3*7}$ &84& 42 & 21 & 21 &1.5382& 1.9755  \\
		\hline
		$\mathbb{Z}_{3*11}$ &132& 66 & 33 & 33 &1.5382& 1.9846  \\
		\hline
		$\mathbb{Z}_{3*5*7}$ & 420&210 & 105 & 105 &1.5382& 1.9952  \\
		\hline
		$\mathbb{Z}_{3*5*11}$ &660& 330 & 165 & 165 &1.5382& 1.9970  \\
		\hline
		$\mathbb{Z}_{3*5*7*11}$ & 4620&2310 & 1155 & 1155 &1.5382& 1.9996  \\
		\hline
		$\mathbb{Z}_{3*5*7*11*13}$ &60060
		& 30030 & 15015 & 15015 &1.5382& 2.0000  \\
		\hline
		$\mathbb{Z}_{3*5*7*11*13*17}$ &1021020& 510510 & 255255 & 255255 &1.5382& 2.0000  \\
		\hline
	\end{tabular}\\
	\caption{Comparison of the parameters of QCSS when the smallest prime factor of $N$ is $3$.\label{new_table_for_3}}
\end{table}

The results of the proposed systematic construction is compared with the results of the existing constructions in the following section.

%In the following section, we compare our proposed constructions with the previous works.

\section{Comparison With Previous Works}
The main difference of this construction with all the previous constructions is that here we can construct asymptotically optimal and near-optimal QCSS over $\mathbb{Z}_N$ for any $N>3$, whereas previously $N\geq 5$ was only odd integer \cite{zhou2020}, prime \cite{Li19_3,Li19_4} or power of prime \cite{Li19_3}. For comparing the parameters with the exixting results, in Table \ref{new_table_for_3}, $K_{prev}$ and $\rho_{prev}$ denote the previous set size and the previously reported optimality factor, respectively, of the QCSS over $\mathbb{Z}_N$, for some values of $N$, when $N$ has a prime factor $3$. The values of $K_{prev}$ and $\rho_{prev}$ are obtained from \cite{zhou2020}. Compared with \cite{Li19_3}, \cite{Li19_4} and \cite{zhou2020}, the uniqueness of our construction can be listed down as follows:
\begin{enumerate}
	\item We can obtain asymptotically optimal QCSS with more flexible parameters using the proposed framework, as compared to the constructions proposed in \cite{Li19_3}, \cite{Li19_4} and \cite{zhou2020}. For instance, the asymptotically optimal and near-optimal QCSS over $\mathbb{Z}_N$, where $N$ is an even composite number and not a prime power, can only be obtained by the proposed construction till date. The parameters and the corresponding optimality factor of the asymptotically optimal QCSS for some of these $N$ can be seen in Table \ref{tab_composite}. Also, the parameters and the corresponding optimality factors of some near-optimal QCSSs are shown in Table \ref{tab_composite1}, where $N$ is even.
	\item When $N$ has the smallest prime factor $3$, near-optimal aperiodic QCSS can be obtained. Compared to the the results reported in \cite{zhou2020}, the obtained QCSSs display a lower optimality factor. Table \ref{new_table_for_3} compares the parameters for some of these values of $N$ with the QCSS designed in \cite{zhou2020}.

%	\item Although for the case when $N$ is a prime number, previous constructions have reported asymptotically optimal aperiodic QCSS, compared to the previous constructions, our proposed construction can generate QCSS with larger set size and lower optimality factor. In Table \ref{tab_prime} we have listed down the parameters of the asymptotically optimal QCSS, where the alphabet size $3\leq p<100$ is odd prime, as example, to show that our proposed QCSS, have a lower optimality factor compared to the previously reported results.
	
%	\item Our proposed construction can generate asymptotically near-optimal QCSS with more flexible parameters which are not covered by the results proposed in \cite{Li19_4}. For example, the asymptotically optimal QCSS, given in Table \ref{new_table_for_3}, can only be constructed by our construction till date.
\end{enumerate}

%\begin{table}[h!]
%	\caption{The comparison when $N=p^2$, $p$ is prime.\label{tab_prime_power}}
%	\small
%	\renewcommand{\arraystretch}{1.3}
%	\resizebox{\textwidth}{!}{
%		\begin{tabular}{|c|c|c|c|c|c|c|c|}
%			\hline
%			% after \\: \hline or \cline{col1-col2} \cline{col3-col4} ...
%			Alphabet & $M$ & $N$& $N_{prev}$& $K$  &$K_{prev}$& $\rho$ &  $\rho_{\min\_prev}$   \\
%			\hline
%			$\mathbb{Z}_{11*11}$ & 121 & 121 &120& 13310 &14641&  1.0601 &  1.0614 \\
%			\hline
%			$\mathbb{Z}_{13*13}$& 169 & 169&168 & 26364  & 28561& 1.0497  & 1.0507 \\
%			\hline
%			$\mathbb{Z}_{17*17}$ & 289 & 289&280& 78608 &83521 & 1.0370 & 1.0376 \\
%			\hline
%			$\mathbb{Z}_{19*19}$ & 361 & 361&360& 123462 &130321& 1.0328 & 1.0333 \\
%			\hline
%			$\mathbb{Z}_{23*23}$ & 529 & 529&520&  267674  &279841& 1.0267 & 1.0271 \\
%			\hline
%			$\mathbb{Z}_{29*29}$ & 841 & 841&840&  682892  &707281& 1.0209 & 1.0211 \\
%			\hline
%			$\mathbb{Z}_{31*31}$ & 961 & 961&960& 893730  &923521& 1.0195 & 1.0197 \\
%			\hline
%			$\mathbb{Z}_{37*37}$ & 1369 & 1369&1360& 1823508 &1874161& 1.0162 & 1.0164 \\
%			\hline
%			$\mathbb{Z}_{41*41}$ & 1681 & 1681&1680& 2756840  &2825761& 1.0146 & 1.0147\\
%			\hline
%			$\mathbb{Z}_{43*43}$ & 1849& 1849&1848& 3339294  &3418801& 1.0139 &1.0140\\
%			\hline
%			$\mathbb{Z}_{47*47}$ & 2209 & 2209&2208& 4775858 &4879681 & 1.0127 & 1.0127\\
%			\hline
%	\end{tabular}}
%\end{table}

\section{Conclusion}
In this paper, we have presented a systematic construction of QCSSs with new flexible parameters based on Florentine rectangles. In the proposed construction, $N>3$ can take any positive value. This construction not only fills the gap left by all the previous constructions, where $N$ was considered odd, prime or prime powers, but also improves the optimality factor as compared with the previous constructions. We first proposed a new set of permutations on $\mathbb{Z}_N$ based on Florentine rectangles and utilized those permutations to construct $(N,N,N)$- CCCs. Combining the newly constructed CCCs we have designed new sets of $(N\times F(N),N,N,N)$- QCSS, where $F(N)$ is the maximum number of rows for which an $F(N)\times N$ Florentine rectangle exists. The designed QCSSs are asymptotically optimal and near-optimal with respect to the correlation bound in \cite{zilong14_1}. For $N=2$ and $N=3$, our construction results to near-optimal $(2N,N,N,N)$- QCSS, with respect to the Welch bound \cite{welch}. We have also compared our proposed construction with the previous constructions reported in the literature. The proposed construction results to new QCSSs when over $\mathbb{Z}_N$, when $N>3$ is an even integer. When $N>3$ is an odd integer, the set size increases and eventually the value of the optimality factor decreases, as compared to the previous constructions.

% We also consider the cases when $N=2\text{ and }3$. In this case we get $(2N,N,N,N)$- QCSS, which are near-optimal with respect to the Welch bound in \cite{welch}. Finally, we compare our proposed work with the previous works and show that the parameters of the proposed QCSS are much more flexible as compared to the parameters of the existing QCSSs.

%Although compared to the systematic constructions, the value of $F(N)$ is higher if we use the computer search results, in this paper we have used the value of $F(N)$, which we have derived from the systematic constructions.

\section*{Appendix A\\Proof of Theorem \ref{theorem2}}

	First, let us prove that, for $0\leq k<F(N)$, $\mathfrak{C}^k$ is an $(N,N,N)$- CCC. Let $\mathcal{C}^{\left(k, m_{1}\right)}, \mathcal{C}^{\left(k, m_{2}\right)} \in \mathfrak{C}^{k},$ where $0\leq k < F(N)$, $0 \leq m_{1}, m_{2} \leq N-1$ and $f_{s}^{(k,m)}(t)$ is as given in (\ref{eq_new_const2}). Then
	
%	Then the aperiodic correlation of  $\mathcal{C}^{\left(k, m_{1}\right)}$ and $\mathcal{C}^{\left(k, m_{2}\right)}$ is

\begin{equation}
\begin{split}
\sum_{s=0}^{N-1} & \tilde{R}_{C_{s}^{\left(k, m_{1}\right)}, C_{s}^{\left(k, m_{2}\right)}}(\tau)\\& =\sum_{s=0}^{N-1} \sum_{t=0}^{N-1-\tau} C_{s, t}^{\left(k, m_{1}\right)} \cdot\left(C_{s, t+\tau}^{\left(k, m_{2}\right)}\right)^{*} \\
& =\sum_{s=0}^{N-1} \sum_{t=0}^{N-1-\tau} \omega_{N}^{f_{s}^{\left(k, m_{1}\right)}(t)} \cdot {\omega_{N}}^{-f_{s}^{\left(k, m_{2}\right)}(t+\tau)}\\
& =\sum_{s=0}^{N-1} \sum_{t=0}^{N-1-\tau} \omega_{N}^{ s(\pi_k(t)-\pi_k(t+\tau))+t\left(m_{1}-m_{2}\right)-m_{2} \tau}.
\end{split}
\end{equation}

consider the following cases.

Case 1: When $\tau=0,m_1=m_2$, then
\begin{equation}
\sum_{s=0}^{N-1} \tilde{R}_{C_{s}^{\left(k, m_{1}\right)}, C_{s}^{\left(k, m_{2}\right)}}(0)=N^2.
\end{equation}

Case 2: When $1\leq\tau\leq N-1,m_1=m_2$,
\begin{equation}\label{eq22}
\begin{split}
\sum_{s=0}^{N-1}& \tilde{R}_{C_{s}^{\left(k, m_{1}\right)},{ C_{s}^{\left(k, m_{2}\right)}}}(\tau) \\& =\sum_{s=0}^{N-1} \sum_{t=0}^{N-1-\tau} \omega_{N}^{-m_{2} \tau} \cdot \omega_{N}^{ s(\pi_k(t)-\pi_k(t+\tau))} \\
& =\omega_{N}^{-m_{2} \tau} \cdot \sum_{t=0}^{N-1-\tau} \sum_{s=0}^{N-1} \omega_{N}^{s(\pi_k(t)-\pi_k(t+\tau))}=0.
\end{split}
\end{equation}
When $\tau \neq 0$, $\pi_k(t)\neq \pi_k(t+\tau)$ because $\pi_k(t)$ is a permutation on $\mathbb{Z}_N$. Also $N\nmid (\pi_k(t)-\pi_k(t+\tau))$ because $(\pi_k(t)-\pi_k(t+\tau)<N$. Therefore (\ref{eq22}) holds.\\

Case 3: When $\tau=0,m_1\neq m_2$,
\begin{equation}
\sum_{s=0}^{N-1} \tilde{R}_{C_{s}^{\left(k, m_{1}\right)},{C_{s}^{\left(k, m_{2}\right)}}}(0)=\sum_{s=0}^{N-1} \sum_{t=0}^{N-1} \omega_{N}^{t \left(m_{1}-m_{2}\right)}=0.
\end{equation}

Case 4: When $1\leq\tau\leq N-1,m_1\neq m_2$,
\begin{equation}\label{neq30}
\begin{split}
\sum_{s=0}^{N-1} &\tilde{R}_{C_{s}^{\left(k, m_{1}\right)},{ C_{s}^{\left(k, m_{2}\right)}}}(\tau)\\&=\sum_{t=0}^{N-1-\tau} \omega_{N}^{t \cdot\left(m_{1}-m_{2}\right)-m_{2} \tau} \cdot \sum_{s=0}^{N-1}\omega_{N}^{s(\pi_k(t)-\pi_k(t+\tau))}=0.
\end{split}
\end{equation}
For $\tau \neq 0$, $\pi_k(t)\neq \pi_k(t+\tau)$, beause $\pi_k(t)$ is a permutation on $\mathbb{Z}_N$. Also $N\nmid (\pi_k(t)-\pi_k(t+\tau))$ because $(\pi_k(t)-\pi_k(t+\tau)<N$. Therefore (\ref{neq30}) holds.

From the above four cases, we conclude that $\mathfrak{C}^{k}$, for each $0\leq k<F(N)$, is an $(N,N,N)$-CCC.

Now let us prove the second part that $\mathfrak{C}$ is a QCSS.
Consider $\mathcal{C}^{\left(k_{1}, m_{1}\right)} \in \mathfrak{C}^{k_1}$ and $\mathcal{C}^{\left(k_{2}, m_{2}\right)} \in \mathfrak{C}^{k_2}$. Then, the ACF of $\mathcal{C}^{\left(k_{1}, m_{1}\right)} $ and  $\mathcal{C}^{(k_{2}, m_{2})}$ is given by
\begin{equation}
\begin{split}
\sum_{s=0}^{N-1} &\tilde{R}_{C_{s}^{\left(k_{1}, m_{1}\right)},{C_{s}^{\left(k_{2}, m_{2}\right)}}}(\tau) \\& =\sum_{s=0}^{N-1} \sum_{t=0}^{N-1-\tau} \omega_{N}^{f_{s}^{\left(k_{1}, m_{1}\right)}(t)} \cdot \omega_{N}^{-f_{s}^{\left(k_{2}, m_{2}\right)}(t+\tau)} \\
& =\sum_{s=0}^{N-1} \sum_{t=0}^{N-1-\tau} \omega_{N}^{t\left(m_{1}-m_{2}\right)-m_{2} \tau+s\left( \pi_{k_{2}}(t+\tau)- \pi_{k_{1}}(t)\right)}\\
& =\sum_{t=0}^{N-1-\tau} \omega_{N}^{t\left(m_{1}-m_{2}\right)-m_{2} \tau} \cdot \sum_{s=0}^{N-1} \omega_{N}^{s\left( \pi_{k_{2}}(t+\tau)- \pi_{k_{1}}(t)\right)}.
\end{split}
\end{equation}

Recall that permutations $\pi_{k_1}$ and $\pi_{k_2}$ satisfy Lemma \ref{lem_new}. Therefore, $ \pi_{k_{1}}(t)- \pi_{k_{2}}(t+\tau)\equiv0\pmod N$ for any $0 \leq t\leq t+\tau \leq N-1, k_1 \neq k_2$ has at most one solution. Hence, if there is no solution, then $\sum\limits_{s=0}^{N-1} \tilde{R}_{C_{s}^{\left(k_{1}, m_{1}\right)}, C_{s}^{\left(k_{2}, m_{2}\right)}}(\tau)=0$ due to $\sum\limits_{s=0}^{N-1} \omega_{N}^{s\left( \pi_{k_{2}}(t+\tau)- \pi_{k_{1}}(t)\right)}=0$. If there is one solution, say $t^\prime$, then for $0\leq t'\leq t^\prime+ \tau\leq N-1$ or in other words for $0\leq t'\leq N-1-\tau$, we have
\begin{equation}
\begin{split}
\sum_{s=0}^{N-1}& \tilde{R}_{C_{s}^{\left(k_{1}, m_{1}\right)},{C_{s}^{\left(k_{2}, m_{2}\right)}}}(\tau)
\\& =\omega_{N}^{-m_{2} \tau} \cdot[\omega_{N}^{(m_{1}-m_{2})\cdot t' }\cdot  N+ \nonumber \\
& \sum_{\substack{0\leq t\leq N-1-\tau, \\ t \neq t'}} \omega_{N}^{\left(m_{1}-m_{2}\right) t}  \sum_{0\leq s\leq N-1} \omega_{N}^{( \pi_{k_{2}}(t+\tau)- \pi_{k_{1}}(t)) \cdot s}] \nonumber\\
& =\omega_{N}^{-m_{2} \tau+\left(m_{1}-m_{2}\right) t^{\prime}} \cdot N.
\end{split}
\end{equation}
Therefore, $|\sum\limits_{s=0}^{N-1} \tilde{R}_{C_{s}^{\left(k_{1}, m_{1}\right)}, {C_{s}^{\left(k_{2}, m_{2}\right)} }}(\tau)| \leq N$ for all $k_1\neq k_2$, $0\leq\tau \leq N-1$ and $0\leq m_1, m_2\leq N-1.$
%%%%%%%%%%%%%%%%%%%%%%%%%%%%%%%%%%%%%%

Therefore, the theorem is proved.

%\begin{IEEEbiography}[{\includegraphics[width=1in,height=1.25in,clip,keepaspectratio]{prof_zhou}}]
%	{Zhengchun Zhou}
%	% or if you just want to reserve a space for a photo:
%	received the B.S. and M.S. degrees in Mathematics
%	and the Ph.D. degree in information security from Southwest Jiaotong
%	University, Chengdu, China, in 2001, 2004, and 2010, respectively.
%	From 2012 to 2013, he was a postdoctoral member in the Department
%	of Computer Science and Engineering, the Hong Kong University
%	of Science and Technology. From 2013 to 2014, he was a research
%	associate in the Department of Computer Science and Engineering,
%	the Hong Kong University of Science and Technology. Since
%	2001, he has been in the Department of Mathematics, Southwest
%	Jiaotong University, where he is currently a professor. His research
%	interests include sequence design, Boolean function, coding theory,
%	and compressed sensing. He is an associated editor of Advances in
%	Mathematics of Communications and IEICE Transactions on Fundamentals,
%	and was a Guest Editor for special issues of Cryptography
%	and Communications.	Dr. Zhou was the recipient of the National excellent Doctoral
%	Dissertation award in 2013 (China).
%\end{IEEEbiography}

\vspace{-0.5in}

%\begin{IEEEbiography}[{\includegraphics[width=1in,height=1.25in,clip,keepaspectratio]{fangrui}}]
%	{Fangrui Liu} was born in Sichuan Province, on April 21, 1996, received the B.S. degree in the School of Mathematics and Computer Science from Quanzhou Normal University, Fujian, China, in 2018. She is currently pursuing the M.S. degree in the School of Mathematics from Southwest Jiaotong University, Chengdu, China. Her research interests include sequence design and its applications.
%\end{IEEEbiography}

\vspace{-0.5in}

%\begin{IEEEbiography}[{\includegraphics[width=1in,height=1.25in,clip,keepaspectratio]{avik}}]
%	{Avik Ranjan Adhikary} is a Postdoctoral fellow at
%	the Department of Mathematics, Southwest Jiaotong
%	University (SWJTU), China. He has received his Ph.D. degree in 2019 from the Department of Mathematics, Indian Institute of Technology Patna (IITP), India. He was a visiting research fellow in the Nanyang Technological University (NTU), Singapore from August 2015 to January
%	2016. He received his M.Sc. degree in Mathematics from the Department of Mathematics,
%	Indian Institute of Technology Guwahati (IITG), Assam, India, in 2013 and his B.Sc. (Hons.) degree in Mathematics from Ramakrishna Mission Vidyamandira (RKMV), Belur Math, Howrah, under University of Calcutta (CU),
%	West Bengal, India, in 2011. He is generally interested in designing sequences with good correlation properties. Details of his research can be found in the
%	link at: https://scholar.google.com/citations?user=y3FBp40AAAAJ\&hl=en
%\end{IEEEbiography}

\vspace{-0.5in}

%\begin{IEEEbiography}[{\includegraphics[width=1in,height=1.25in,clip,keepaspectratio]{prof_fan}}]
%	{Pingzhi Fan (M’93-SM’99-F’15)}
%	% or if you just want to reserve a space for a photo:
%	received his MSc degree in computer science from the Southwest Jiaotong University, China, in 1987, and PhD degree in Electronic Engineering from the Hull University, UK, in 1994. He is currently a distinguished professor and director of the institute of mobile communications, Southwest Jiaotong University, China, and a visiting professor of Leeds University, UK (1997-), a guest professor of Shanghai Jiaotong University (1999-). He is a recipient of the UK ORS Award (1992), the NSFC Outstanding Young Scientist Award (1998), IEEE VTS Jack Neubauer Memorial Award (2018), and 2018 IEEE SPS Signal Processing Letters Best Paper Award. His current research interests include vehicular communications, massive multiple access and coding techniques, etc. He served as general chair or TPC chair of a number of international conferences including VTC’2016Spring, IWSDA’2019, ITW’2018 etc. He is the founding chair of IEEE Chengdu (CD) Section, IEEE VTS BJ Chapter and IEEE ComSoc CD Chapter. He also served as an EXCOM member of IEEE Region 10, IET(IEE) Council and IET Asia Pacific Region. He has published over 300 international journal papers and 8 books (incl. edited), and is the inventor of 25 granted patents. He is an IEEE VTS Distinguished Lecturer (2015-2019), and a fellow of IEEE, IET, CIE and CIC.
%\end{IEEEbiography}

\end{document}